\documentclass[11pt]{article}
\usepackage[T1]{fontenc}

\usepackage{stix2}
\usepackage{a4wide}
\usepackage{authblk}

\usepackage{amsmath}

\usepackage{graphicx}
\usepackage{subcaption}

\usepackage{pgf}
\usepackage{import}

\usepackage{tikz}
\usetikzlibrary{calc,decorations.pathmorphing,decorations.pathreplacing}
\usepackage{hyperref}
\usepackage{cleveref}
\usepackage{sectsty}
\allsectionsfont{\raggedright}

\title{Pseudo-spectral frequency-domain method with background field decomposition and Green's function preconditioner for electromagnetic scattering problem in EUV lithography}
\author{Seungjin Lee\thanks{seungjin.lee@imec.be}}
\author{Werner Gillijns\thanks{werner.gillijns@imec.be}}
\author{Doyun Kim\thanks{doyun.kim@imec.be}}
\affil{IMEC, Kapeldreef 75, 3001 Leuven, Belgium}
\date{}

\begin{document}

\maketitle
\begin{abstract}
    We provide an accelerated computational framework to solve electromagnetic scattering problems in planarly layered media arising from extreme ultraviolet (EUV) lithography. To achieve this, we reformulate the EUV scattering problem into a scattering problem on a homogeneous background, in which the electromagnetic contribution of the layered media is captured by a recursively updated reflection of the layered stack. The system is numerically solved by employing the pseudo-spectral frequency-domain method paired with an iterative solver, whose iterative convergence is expedited by a free-space Green's function preconditioner. The proposed framework is evaluated on EUV mask geometries and multilayer mirror stacks, demonstrating a significant speedup over the conventional pseudo-spectral frequency-domain method.
\end{abstract} 

\section{Introduction}
\label{sec:introduction}

Accurate and computationally efficient electromagnetic simulation is a fundamental challenge in computational science, dictated by the ubiquity of light-matter interactions across academic research and industrial domains. Among the diverse applications driving this challenge, optical lithography is of particular interest due to the intense industrial demand for the precise and accelerated simulation of the optical components within a lithography process.

Traditionally, the optical component of lithography simulation has been modeled using two-dimensional Kirchhoff diffraction, often termed the thin-mask approximation \cite{wong2001,mack2007,ma2011}. This simplified approach has been sufficient for legacy technology nodes, where the physical features on a photomask remain large enough for scalar and paraxial wave approximations to hold. Furthermore, any localized inaccuracies of the optical simulation have been effectively compensated by data-driven calibrations designed to account for the complex chemical steps beyond optics. Finally, lithographic success has been primarily evaluated using low-dimensional metrics such as the critical dimension (CD), which have been inadequate to reveal the fundamental shortcomings of the simplified optics model.

However, the emergence to EUV lithography pushes the underlying optical physics far beyond the scope of traditional approximations. Crucially, the EUV source of wavelength $13.5$ nm coupled to oblique illumination angles leads to the reflective optics with strong three-dimensional vector diffraction effects, which  invalidates conventional scalar and paraxial approximations \cite{erdmann2017, erdmann2019}. Also, the stringent demand for sub-nanoscale wafer features have prompted the introduction of more sophisticated measurement metrics to evaluate complex three-dimensional wafer structures, which exhibit an acute sensitivity to the details of the underlying electromagnetic field distribution \cite{levinson2020}.

This profound technical shift to EUV lithography mandates that the underlying simulation framework incorporates rigorous electromagnetic field solvers. Conventional Maxwell solvers, such as the finite-difference time-domain (FDTD) method \cite{yee1966,taflove1998} and rigorous coupled-wave analysis (RCWA) \cite{moharam1981, moharam1995}, simulate complex light-mask interactions with high fidelity. However, despite their accuracy, the substantial computational cost of these traditional Maxwell solvers remains incompatible with high-throughput industrial applications like Inverse Lithography Technology (ILT) and iterative process optimization. 

To bridge the gap between accuracy and computational throughput, we present an accelerated numerical method to solve a class of electromagnetic scattering problems arising from EUV lithography. The core efficiency of the proposed method is rooted in a pseudo-spectral frequency-domain (PSFD) formulation \cite{liu1997} tailored for solving time-harmonic electromagnetic scattering problems on planarly layered media. By employing a Fourier basis coupled with a uniform collocation grid, we construct a discrete numerical system whose forward pass can be efficiently computed via the fast Fourier transform (FFT) while preserving spectral accuracy.

To further mitigate the computational overhead inherent to the structural complexity of the problem, we introduce a background field decomposition \cite{abbott1982} in which the background field corresponds to the response of the layered media to both the incident and the secondary fields scattered by the scattering potential on top of the layered media. The decomposition fully leverages the geometrical configuration of the problem, where the scattering potential varying across the transverse plane is strictly on top of the layered media stack. In turn, the scattering problem can be formulated as a free-space scattering problem with a source term derived from the background field, yielding a fixed computational domain bounded around the top scattering potential independent of the thickness and the complexity of the underlying layered media.

The remaining free-space scattering problem is solved using an iterative solver, whose convergence is expedited by employing a free-space Green's function as a preconditioner. The action of the preconditioner maps the scattering problem into the well-known Lippmann-Schwinger formulation \cite{chew1995,chew2001,colton2020} identical to the problem defined by the time-harmonic Maxwell's equations in homogeneous space. Once coupled with the bounded scattering potential, the Green's function preconditioner achieves a drastic enhancement of the convergence, which allows us to interprete the resulting Lippmann-Schwinger formulation as a well-preconditioned reformulation of the scattering problem based on the Maxwell's equations.

This paper is organized as follows. In \Cref{sec:problem}, we provide a mathematical formulation of the electromagnetic scattering problem arising from EUV lithography. This includes the geometry of the scattering potential of the problem, together with an explicit exposition of the background field decomposition. In \Cref{sec:num_psfd}, we present our PSFD method for solving the problem formulated in \Cref{sec:problem} by providing the full implementation details. We also provide numerical demonstrations of the solver routine for typical configurations of the EUV scattering problem. Finally, \Cref{sec:conclusion} is devoted to a summary of the paper and an outlook on future research directions.

\section{Electromagnetic scattering problem for EUV lithography}
\label{sec:problem}

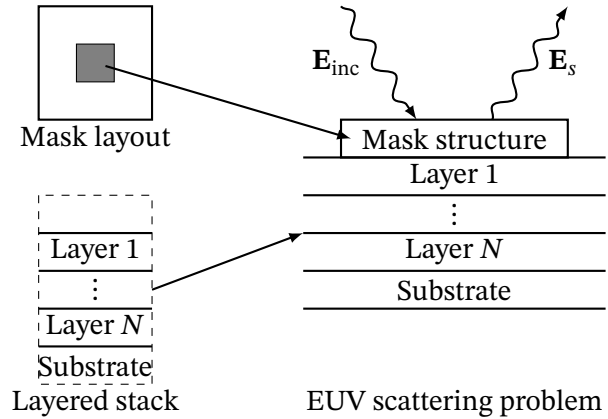
\begin{figure}[htbp]
    \centering
    \begin{tikzpicture}
        \draw[thick] (0.0,0.0) rectangle (1.5,-1.5) node[pos=0.5, yshift=-1.0cm] (mask layout) {Mask layout};
        \draw[fill = black!50] (0.5,-0.5) rectangle (1,-1) node[pos=0.5] (absorber layer) {};
        \draw[dashed] (0.0,-2.5) rectangle (1.5,-5) node[pos=0.5, yshift=-1.5cm] {Layered stack};
        \draw[thick] (0.0,-3) -- (1.5,-3) node[pos=0.5, yshift=-0.25cm] {Layer $1$}; ;
        \draw[thick] (0.0,-3.5) -- (1.5,-3.5) node[pos=0.5, yshift=-0.25cm] (center) {\vdots}; ;
        \draw[thick] (0.0,-4) -- (1.5,-4) node[pos=0.5, yshift=-0.25cm] {Layer $N$}; ;
        \draw[thick] (0.0,-4.5) -- (1.5,-4.5) node[pos=0.5, yshift=-0.25cm] {Substrate}; ;

        \draw[-latex, decoration={snake, amplitude=1mm, segment length=5mm}, decorate, thick] (4,0) -- node[midway, left, xshift=-0.1cm] {$\mathbf{E}_{\mathrm{inc}}$} (5.,-1.5);
        \draw[-latex, decoration={snake, amplitude=1mm, segment length=5mm}, decorate, thick] (6,-1.5) -- node[midway, right, xshift=0.1cm] {$\mathbf{E}_{s}$} (7,0);
        \draw[thick] (4,-1.5) rectangle (7,-2) node[pos=0.5] (mask structure) {Mask structure};
        \draw[thick] (3.5,-2) -- (7.5,-2);
        \draw[thick] (3.5,-2.5) -- (7.5,-2.5);
        \draw[thick] (3.5,-3) -- (7.5,-3);
        \draw[thick] (3.5,-3.5) -- (7.5,-3.5);
        \draw[thick] (3.5,-4) -- (7.5,-4);
        \node at (5.5,-2.25) {Layer 1};
        \node at (5.5,-2.75) {\vdots};
        \node at (5.5,-3.25) {Layer $N$};
        \node at (5.5,-3.75) {Substrate};
        \node at (0,-5) {};
        \draw[-latex, thick] (absorber layer) -- (mask structure.west);
        \draw[-latex, thick] (1.5, -3.75) -- (3.5, -3);
        \node at (5.5, -5.25) {EUV scattering problem};
    \end{tikzpicture}
    \caption{
   Schematic representation of the EUV scattering problem. The scattering potential is defined by a two-dimensional mask layout and a layered stack. Interaction of the incident field $\mathbf{E}_{\mathrm{inc}}$ with the mask structures and underlying thin-film layers generates the resulting total scattered field $\mathbf{E}_{s}$.
   }
    \label{fig:euv_scattering_problem}
\end{figure}

The scattering problem of EUV lithography can be modeled by a time-harmonic electromagnetic scattering problem on a planarly layered media, in which a spatially varying scattering potential is situated atop of the layered media. The multilayer stack typically consists of alternating layers of molybdenum (Mo) and silicon (Si) that serves as a Bragg reflector engineered to reflect incident light at the EUV wavelength. Conversely, the top scattering potential is designed to modulate the reflected field by absorbing the incident light. The scattering potential on top of the layered media is vertically confined within a thin layer, called the absorber layer, and its distribution across the transverse plane is completely determined by a two-dimensional layout termed the mask layout. A schematic description of the geometric configuration is provided in \Cref{fig:euv_scattering_problem}.

Assuming a homogeneous magnetic permeability $\mu$, the EUV scattering problem can be formulated from the time-harmonic Maxwell's equations, yielding the vector Helmholtz equation for the scattered electric field $\mathbf{E}_{s}$ \cite{born1999,jackson2009}:
\begin{align}
    \nabla \times \nabla \times \mathbf{E}_{s}(\mathbf{r}) - \omega^2 \varepsilon(\mathbf{r}) \mu \mathbf{E}_{s}(\mathbf{r}) = \omega^2 \Delta \varepsilon(\mathbf{r}) \mu \mathbf{E}_{\mathrm{inc}}(\mathbf{r}), \quad \Delta \varepsilon(\mathbf{r}) = \varepsilon(\mathbf{r}) - \varepsilon_{\mathrm{bg}},
    \label{eq:euv_scattering_problem}
\end{align}
where $\varepsilon(\mathbf{r})$ and $\varepsilon_{\mathrm{bg}}$ represent the total and background permittivities, respectively. The incident field $\mathbf{E}_{\mathrm{inc}}$ satisfies the homogeneous vector Helmholtz equation dictated by the background permittivity $\varepsilon_{\mathrm{bg}}$, and $\omega$ denotes the angular frequency. The vector Helmholtz equation in \eqref{eq:euv_scattering_problem} is accompanied by an asymptotic radiation condition, such as the Silver-Müller radiation condition, to reflect the asymptotic behavior at infinity.

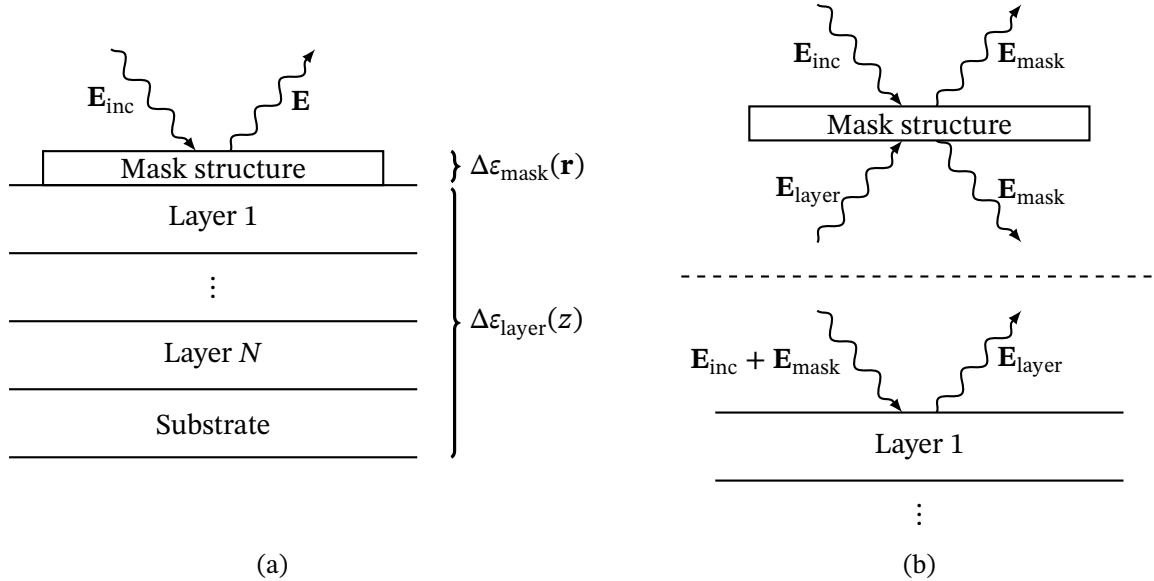
\begin{figure}[htbp]
    \centering
    \begin{subfigure}[t]{0.45\textwidth}
        \centering
    \begin{tikzpicture}[scale=0.9]
        \draw[-latex, decoration={snake, amplitude=1mm, segment length=5mm}, decorate, thick] (-1.5,2.0) -- node[midway, left, xshift=-0.1cm] {$\mathbf{E}_{\mathrm{inc}}$} (-0.25,0.5);
        \draw[-latex, decoration={snake, amplitude=1mm, segment length=5mm}, decorate, thick] (0.25,0.5) -- node[midway, right, xshift=0.1cm] {$\mathbf{E}$} (1.5,2.0);
        \draw[thick] (-3,0) -- (3,0);
        \draw[thick] (-3,-1) -- (3,-1);
        \draw[thick] (-3,-2) -- (3,-2);
        \draw[thick] (-3,-3) -- (3,-3);
        \draw[thick] (-3,-4) -- (3,-4);
        \draw[thick] (-2.5,0.5) rectangle (2.5,0);
        \node at (0,0.25) {Mask structure};
        \node at (0,-0.5) {Layer 1};
        \node at (0,-1.5) {\vdots};
        \node at (0,-2.5) {Layer $N$};
        \node at (0,-3.5) {Substrate};
        \draw [very thick, decorate, decoration = {brace}] (3.5,0.5) --  (3.5, 0.05) node [midway, right, xshift=0.1cm] {$\Delta \varepsilon_{\mathrm{mask}}(\mathbf{r})$};
        \draw [very thick, decorate, decoration = {brace}] (3.5,-0.05) --  (3.5,-4) node [midway, right, xshift=0.1cm] {$\Delta \varepsilon_{\mathrm{layer}}(z)$};
        \node at (0,-5) {};
    \end{tikzpicture}
        \caption{}
        \label{fig:euv_scattering_problem_full}
    \end{subfigure}
    \hfill 
    \begin{subfigure}[t]{0.45\textwidth}
        \centering
    \begin{tikzpicture}[scale=0.9]
        \draw[-latex, decoration={snake, amplitude=1mm, segment length=5mm}, decorate, thick] (-1.5,2.0) -- node[midway, left, xshift=-0.1cm] {$\mathbf{E}_{\mathrm{inc}}$} (-0.25,0.5);
        \draw[-latex, decoration={snake, amplitude=1mm, segment length=5mm}, decorate, thick] (-1.5,-1.5) -- node[midway, left, xshift=-0.1cm] {$\mathbf{E}_{\mathrm{layer}}$} (-0.25,0.0);
        \draw[-latex, decoration={snake, amplitude=1mm, segment length=5mm}, decorate, thick] (0.25,0.5) -- node[midway, right, xshift=0.1cm] {$\mathbf{E}_{\mathrm{mask}}$} (1.5,2.0);
        \draw[-latex, decoration={snake, amplitude=1mm, segment length=5mm}, decorate, thick] (0.25,0.0) -- node[midway, right, xshift=0.1cm] {$\mathbf{E}_{\mathrm{mask}}$} (1.5,-1.5);
        \node at (0,0.25) {Mask structure};
        \draw[thick] (-2.5,0.5) rectangle (2.5,0);
        \draw[dashed, thick] (-3.5,-2) -- (3.5,-2);
        \draw[-latex, decoration={snake, amplitude=1mm, segment length=5mm}, decorate, thick] (-1.5,-2.5) -- node[midway, left, xshift=-0.1cm] {$\mathbf{E}_{\mathrm{inc}} + \mathbf{E}_{\mathrm{mask}}$} (-0.25,-4);
        \draw[-latex, decoration={snake, amplitude=1mm, segment length=5mm}, decorate, thick] (0.25,-4) -- node[midway, right, xshift=0.1cm] {$\mathbf{E}_{\mathrm{layer}}$} (1.5,-2.5);
        \draw[thick] (-3,-4) -- (3,-4);
        \draw[thick] (-3,-5) -- (3,-5);
        \node at (0,-4.5) {Layer 1};
        \node at (0,-5.5) {\vdots};
    \end{tikzpicture}
        \caption{}
        \label{fig:euv_scattering_problem_decomposed}
    \end{subfigure}
    \caption{
        Decomposition of the EUV scattering problem via the background field framework. (a) Total structural layout of the mask and substrate. (b) Equivalent split representation decoupled into an isolated mask scattering problem and an analytical multi-layer boundary response.
        }
    \label{fig:euv_scattering_problem_decomposition}
\end{figure}

By noticing that the mask structure and the layered media are spatially disjoint, it is computationally advantageous to separate their individual electromagnetic contributions, so that one can solve for the scattered field across the mask region perturbed on a background field derived from the stratified media. This separation can be achieved by employing the classical background field method \cite{abbott1982}, where the total scattered field $\mathbf{E}_{s}$ in \eqref{eq:euv_scattering_problem} is decomposed into a background field $\mathbf{E}_{\mathrm{layer}}$---which satisfies Maxwell's equations in the layered media in the absence of the mask---and a scattered field $\mathbf{E}_{\mathrm{mask}}$ induced by the presence of the mask structures. Consequently, the vector Helmholtz equation in \eqref{eq:euv_scattering_problem} can be decomposed into two coupled equations for $\mathbf{E}_{\mathrm{mask}}$ and $\mathbf{E}_{\mathrm{layer}}$, as schematically illustrated in \Cref{fig:euv_scattering_problem_decomposition}:
\begin{subequations}
\begin{alignat}{3}
    &\nabla \times \nabla \times \mathbf{E}_{\mathrm{mask}}(\mathbf{r}) &&- \omega^2 (\varepsilon_{\mathrm{bg}} + \Delta \varepsilon_{\mathrm{mask}}(\mathbf{r})) \mathbf{E}_{\mathrm{mask}}(\mathbf{r}) &&= \omega^2 \Delta \varepsilon_{\mathrm{mask}}(\mathbf{r}) (\mathbf{E}_{\mathrm{inc}}(\mathbf{r}) + \mathbf{E}_{\mathrm{layer}}(\mathbf{r})),\\
    &\nabla \times \nabla \times \mathbf{E}_{\mathrm{layer}}(\mathbf{r}) &&- \omega^2 (\varepsilon_{\mathrm{bg}} + \Delta \varepsilon_{\mathrm{layer}}(\mathbf{r})) \mathbf{E}_{\mathrm{layer}}(\mathbf{r}) &&= \omega^2 \Delta \varepsilon_{\mathrm{layer}}(\mathbf{r}) (\mathbf{E}_{\mathrm{inc}}(\mathbf{r}) + \mathbf{E}_{\mathrm{mask}}(\mathbf{r})).
\end{alignat}
\label{eq:euv_scattering_problem_decomposed}
\end{subequations}

The spatial disjointness then allows the background field $\mathbf{E}_{\mathrm{layer}}$ to be identified as the  response of the layered media to both the incident field $\mathbf{E}_{\mathrm{inc}}$ and the secondary field $\mathbf{E}_{\mathrm{mask}}$ scattered downward from the mask topography \cite{chew1995,chew1985}. In turn, the system reduces to a scattering problem for $\mathbf{E}_{\mathrm{mask}}$, driven by a self-consistent source term that implicitly accounts for substrate reflections:
\begin{align}
    \nabla \times \nabla \times \mathbf{E}_{\mathrm{mask}} - \omega^2 \left(\varepsilon_{\mathrm{bg}} + \Delta \varepsilon_{\mathrm{mask}}\right) \mathbf{E}_{\mathrm{mask}} = \omega^2 \Delta \varepsilon_{\mathrm{mask}} \left(\mathbf{E}_{\mathrm{inc}} + \mathcal{R}\left[\mathbf{E}_{\mathrm{inc}} + \mathbf{E}_{\mathrm{mask}}\right]\right),
    \label{eq:field_recursive}
\end{align}
where $\mathcal{R}$ denotes the generalized reflection operator of the layered media.

In solving the scattering problem \eqref{eq:field_recursive}, the electromagnetic response of the layered stack can be analytically computed by employing the transfer matrix method \cite{born1999, mackay2020} or the generalized reflection coefficients \cite{chew1995}. This allows the computational domain to be restricted to the bounded region surrounding the mask structure on a homogeneous background without including the entire layered media. As demonstrated in \Cref{sec:num_psfd}, excluding the layered stack for the active computational domain drastically lowers the numerical complexity of the scattering problem with a large number of layers. 

The background field method is closely related to the Fourier boundary-condition method used in time-domain formulations, such as FDTD \cite{pistor1998}, where the response of the multilayer stack provides a boundary condition for the scattered field at the interface between the mask structure and the stratified substrate. However, the free-space scattering problem in \eqref{eq:field_recursive} offers a computational advantage over the conventional domain decomposition methods since an iterative solver routine can be accelerated using a free-space Green's function as a preconditioner. This preconditioning strategy is generally not applicable to standard domain decomposition methods, where the solution field is subject to non-trivial boundary conditions derived from the layered media.

\section{Pseudo-spectral frequency-domain method and free-space Green's function preconditioner} 
\label{sec:num_psfd}
To solve the scattering problem formulated in \eqref{eq:field_recursive}, we employ a pseudo-spectral method using a finite Fourier series as the computational basis \cite{boyd2001,fornberg1998}. Formally, the pseudo-spectral approach approximates the solution $u(x)$ of the differential equation as:
\begin{align}
   u(x) \approx \sum_{n=1}^{N} \widehat{u}_n e^{i k_n x},
   \label{eq:fourier_basis}
\end{align}
where $\widehat{u}_n$ are the expansion coefficients and $k_n$ are the corresponding discrete wavenumbers. The action of a linear operator $\mathcal{L}$ on $u(x)$ can then be expressed as:
\begin{align}
   \mathcal{L} u(x) = \sum_{n=1}^{N} \widehat{u}_n (\mathcal{L} e^{i k_n x}).
   \label{eq:fourier_basis_diff}
\end{align}
By evaluating the expansion \eqref{eq:fourier_basis_diff} at a set of collocation points, the operator $\mathcal{L}$ is mapped to a finite matrix acting on the spectral coefficient vector $\mathbf{\widehat{u}}$.

The PSFD method \cite{liu1997} exploits a uniform collocation grid to implement the pseudo-spectral formulation in a Fourier basis, which enables the efficient computation of expansion coefficients via the Fast Fourier Transform (FFT). Explicitly, by adopting a uniform spatial grid composed of $\left( N_{x}, N_{y}, N_{z} \right)$ sampling points, the unknown vector field $\mathbf{E}_{\mathrm{mask}}(\mathbf{r})$ in \eqref{eq:field_recursive} maps to a $(3N_x N_y N_z)$-dimensional complex algebraic vector. Correspondingly, the continuous linear operator $\nabla \times \nabla \times - \, \omega^{2} \big(\varepsilon_{\mathrm{bg}} + \Delta \varepsilon_{\mathrm{mask}}(\mathbf{r})\big)$ translates to a discrete operator matrix representation of size $(3N_x N_y N_z) \times (3N_x N_y N_z)$. Rather than explicitly assembling this dense system, the spatial derivatives are evaluated matrix-free in the spectral domain via the FFT:
\begin{align}
    \partial_{\alpha} \mapsto \mathrm{FFT}_{\alpha}^{-1} \left( i k_{\alpha} \, \mathrm{FFT}_{\alpha}[\cdot] \right), \quad \alpha \in \{x, y, z\},
    \label{eq:spectral_derivative}
\end{align}
where $\mathrm{FFT}_{\alpha}$ denotes the one-dimensional FFT executed along the ${\alpha}$-direction, and $k_{\alpha}$ represents the corresponding discrete wavenumber array.

\begin{figure}[t]
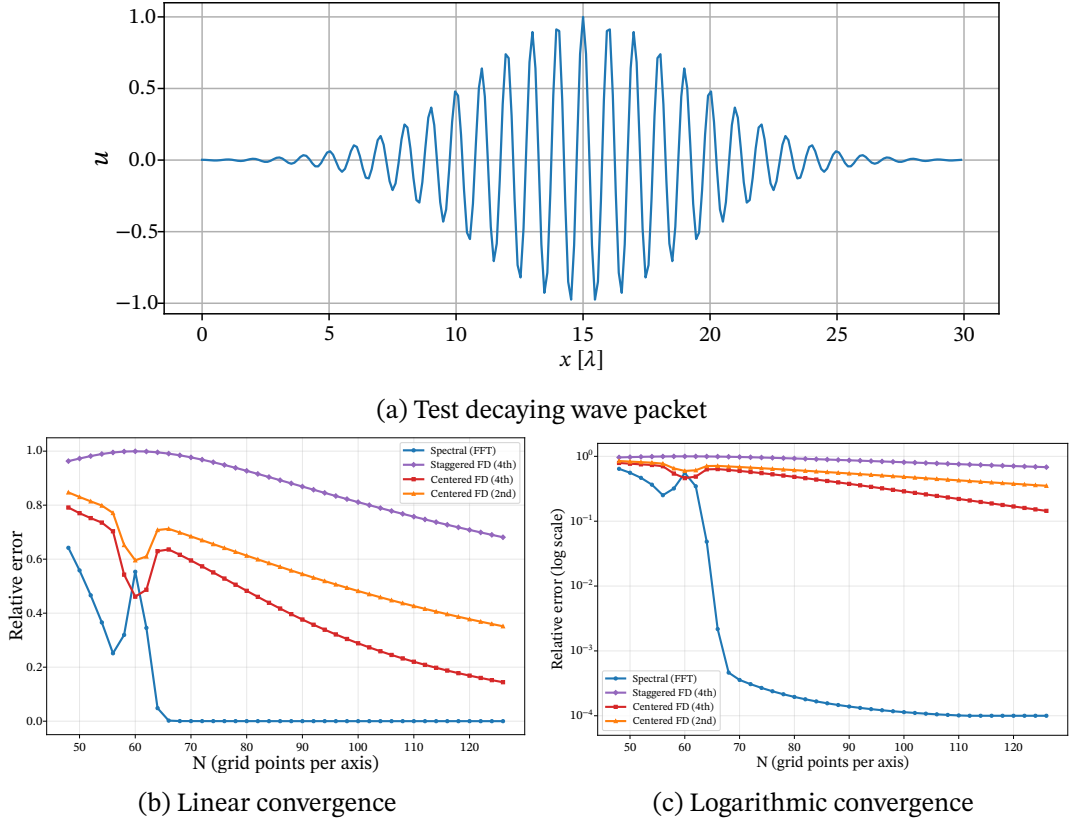

    \centering
    \subcaptionbox{Test decaying wave packet\label{fig:spectral_convergence_wave}}{%
        \resizebox{0.8\textwidth}{!}{%
        \let\oldfontsize\fontsize
        \renewcommand{\fontsize}[2]{%
            \oldfontsize{\dimexpr1.3\dimexpr#1pt\relax\relax}{\dimexpr1.3\dimexpr#2pt\relax\relax}%
            \selectfont
        }
            \input{curl_curl_convergence_wave.pgf}%
        \let\fontsize\oldfontsize
        }%
    }
    \subcaptionbox{Linear convergence \label{fig:spectral_convergence_linear}}{%
        \resizebox{0.45\textwidth}{!}{%
            \input{curl_curl_convergence_linear.pgf}
        }
    }
    \subcaptionbox{Logarithmic convergence\label{fig:spectral_convergence_log}}{%
        \resizebox{0.45\textwidth}{!}{%
            \input{curl_curl_convergence_log.pgf}
        }
    }
    \caption{
        Numerical convergence analysis for the $\nabla \times \nabla \times$ differential operator. (a) Isotropic decaying wavefield profile $u$ spanning a $30\lambda$ isotropic domain. (b), (c) Convergence rate comparisons showing the exponential accuracy of the proposed PSFD scheme versus the algebraic scaling of the conventional finite-difference method with respect to the grid points $N$ per axis.
        }
    \label{fig:spectral_convergence}
    
\end{figure}

The primary advantage of the pseudo-spectral approach stems from the fact that the truncation error decays exponentially (spectrally) with the number of expansion modes, rather than polynomially (algebraically) with the grid spacing. This spectral convergence offers a numerical benefit over conventional finite-difference schemes, which approximate spatial derivatives using localized stencils bounded by a staggered Yee grid \cite{yee1966}. As an illustrative example, \Cref{fig:spectral_convergence} demonstrates the rapid convergence of the continuous double-curl operator $\nabla \times \nabla \times$ evaluated matrix-free on a smooth, decaying wavefield via the Fourier derivative in \eqref{eq:spectral_derivative}.

The pseudo-spectral method has been applied across a wide range of computational electromagnetics problems, consistently demonstrating high accuracy while using a coarser grid sampling compared to conventional domain-discretization methods such as the finite-difference time-domain (FDTD) method or the finite-element method (FEM) \cite{liu1997,pistor1998,liu1999,song2011,munro2014}. For a comprehensive exposition of the underlying mathematical theory and stability aspects of the pseudo-spectral framework, we refer the reader to standard literature \cite{boyd2001,fornberg1998}.

The spectral representation of the differential operator in \eqref{eq:spectral_derivative} enforces periodic boundary conditions on the computational domain, whereas the radiation condition of the open scattering problem requires waves to propagate outward freely without re-entering the domain boundaries. To satisfy the open radiation condition within the FFT formulation, the bounded domain is enclosed by perfectly matched layers (PMLs) \cite{berenger1994}. A PML acts as an artificial, impedance-matched lossy medium that exponentially attenuates outgoing waves before they reach the periodic boundaries, thereby suppressing wrap-around reflections. The PML can be realized via a complex coordinate stretching that scales the spatial derivative operators to form the modified pseudo-spectral differential operators,
\begin{align}
    \partial_{\alpha} \mapsto \frac{1}{1 + i \tilde{\sigma}_{\alpha}(\alpha)} \, \mathrm{FFT}_{\alpha}^{-1} \left( i k_{\alpha} \, \mathrm{FFT}_{\alpha}[\cdot] \right), \quad \alpha \in \{x, y, z\},
    \label{eq:pml_derivative}
\end{align}
where $\tilde{\sigma}_{\alpha}(\alpha) = \sigma_{\alpha}(\alpha)/(\omega \varepsilon)$ represents the normalized PML conductivity profile defined along the corresponding coordinate axis.

\begin{figure}[htbp]
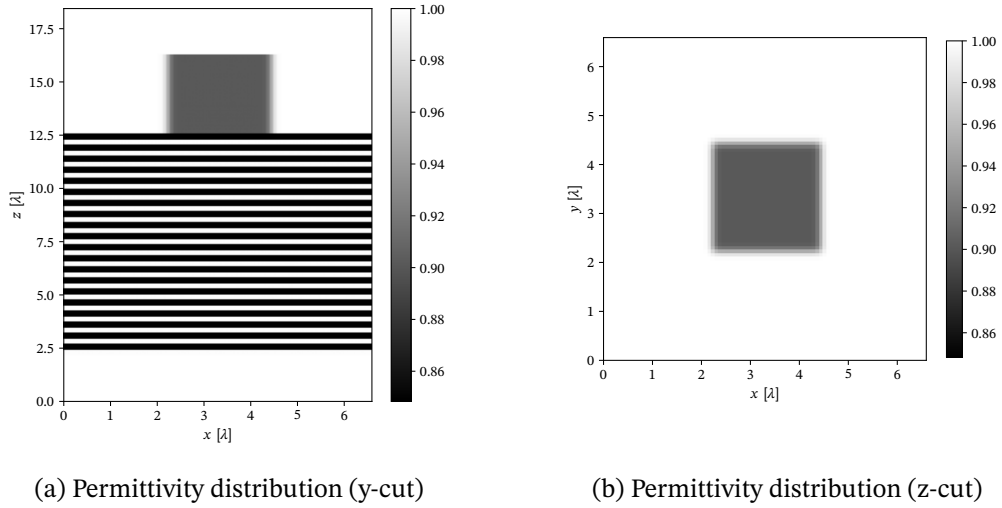

    \centering
    \subcaptionbox{Permittivity distribution (y-cut)\label{fig:cuboid_problem_ycut}}{%
        \resizebox{0.4\textwidth}{!}{%
            \input{eps_ycut.pgf}%
        }%
    }
    \hspace{0.06\textwidth}
    \subcaptionbox{Permittivity distribution (z-cut)\label{fig:cuboid_problem_zcut}}{%
        \resizebox{0.4\textwidth}{!}{%
            \input{eps_zcut.pgf}%
        }%
        \vspace{0.28cm}
    }
    \caption{
        Cross-sectional permittivity maps of the test structure used for verifying the PSFD solver against the reference FDFD method. The setup includes a $50\,$nm-thick Ta absorber mask layer placed on a Mo/Si multilayer Bragg reflector (consisting of $20$ bilayers, each with a $7\,$nm period).
        }
    \label{fig:cuboid_problem}
\end{figure}

\begin{figure}[t]
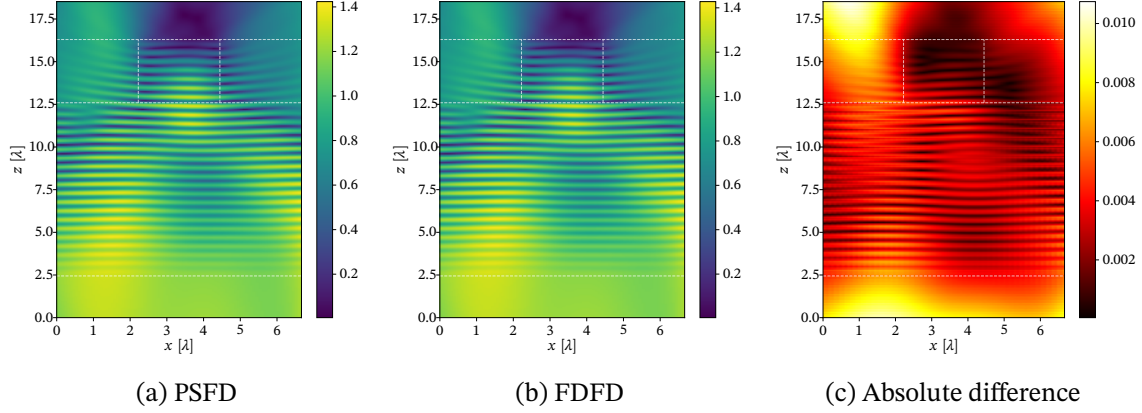

    \centering
    \subcaptionbox{PSFD\label{fig:cuboid_problem_psfd}}{%
        \resizebox{0.32\textwidth}{!}{%
        \let\oldfontsize\fontsize
        \renewcommand{\fontsize}[2]{%
            \oldfontsize{\dimexpr1.3\dimexpr#1pt\relax\relax}{\dimexpr1.3\dimexpr#2pt\relax\relax}%
            \selectfont
        }
            \input{Ey_spectral_Nxy150.pgf}%
        \let\fontsize\oldfontsize
        }%
    }
    %\hspace{0.08\textwidth}
    \subcaptionbox{FDFD\label{fig:cuboid_problem_fdfd}}{%
        \resizebox{0.32\textwidth}{!}{%
        \let\oldfontsize\fontsize
        \renewcommand{\fontsize}[2]{%
            \oldfontsize{\dimexpr1.3\dimexpr#1pt\relax\relax}{\dimexpr1.3\dimexpr#2pt\relax\relax}%
            \selectfont
        }
            \input{Ey_fdfd_Nxy150.pgf}%
        \let\fontsize\oldfontsize
        }%
    }
    \subcaptionbox{Absolute difference\label{fig:cuboid_problem_absdiff}}{%
        \resizebox{0.32\textwidth}{!}{%
        \let\oldfontsize\fontsize
        \renewcommand{\fontsize}[2]{%
            \oldfontsize{\dimexpr1.3\dimexpr#1pt\relax\relax}{\dimexpr1.3\dimexpr#2pt\relax\relax}%
            \selectfont
        }
            \input{Ey_diff_Nxy150.pgf}%
        \let\fontsize\oldfontsize
        }%
    }
    \caption{
    Intensity distributions $\vert E_y\vert^2$ of the scattered electric fields on the $y=0$ plane, computed by (a) PSFD and (b) the FDFD together with (c) their absolute difference. The white dashed lines indicate the interfaces of the Ta absorber and the layered stack.}
    \label{fig:cuboid_problem_solved}
\end{figure}

To evaluate the predictive accuracy and performance of the PSFD method, we benchmark the PSFD method on a simple test problem against the conventional finite-difference frequency-domain (FDFD) method. We consider a configuration modeling a standard EUV mask stack \cite{makhotkin2021}, schematically detailed in \Cref{fig:cuboid_problem}. The model consists of a $50\,$nm-thick tantalum (Ta) absorber cuboid with width $30\,$nm deposited atop a stratified $\mathrm{Mo/Si}$ distributed Bragg reflector comprising $20$ bilayer pairs with a spatial period of $7\,$nm. The system is excited by an TE-polarized plane wave at a wavelength of $13.5\,$nm under a $6^{\circ}$ oblique incidence angle. To enforce open boundary conditions, $30\,$nm-thick perfectly matched layers (PMLs) are deployed along all six faces of the three-dimensional grid space. A comparison of the resulting field solutions is presented in \Cref{fig:cuboid_problem_solved}.

\begin{figure}[h]
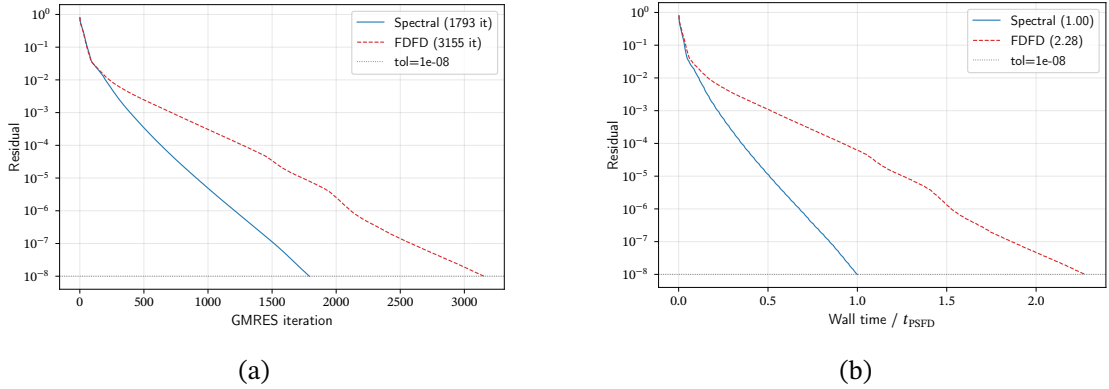

    \centering
    \subcaptionbox{\label{fig:cuboid_problem_convergence_iteration}}{%
        \resizebox{0.45\textwidth}{!}{%
        \let\oldfontsize\fontsize
        \renewcommand{\fontsize}[2]{%
            \oldfontsize{\dimexpr1.0\dimexpr#1pt\relax\relax}{\dimexpr1.3\dimexpr#2pt\relax\relax}%
            \selectfont
        }
            \input{solver_log_iteration.pgf}%
        \let\fontsize\oldfontsize
        }%
    }
    \hspace{0.05\textwidth}
    \subcaptionbox{\label{fig:cuboid_problem_convergence_walltime}}{%
        \resizebox{0.45\textwidth}{!}{%
            \input{solver_log_walltime.pgf}%
        }%
    }
    \caption{The convergence of the PSFD and FDFD methods for the test problem in \Cref{fig:cuboid_problem}. The PSFD method converges faster than the FDFD method in terms of (a) the number of iterations and (b) the computational time.}
    \label{fig:cuboid_problem_convergence}
\end{figure}

The discrete electric field distributions are obtained by solving the respective linear systems derived from the PSFD and FDFD formulations using the GMRES algorithm \cite{saad1986,saad2003}, configured with a relative residual convergence tolerance of $10^{-8}$. The computed field profiles are highlighted in \Cref{fig:cuboid_problem_psfd,fig:cuboid_problem_fdfd}. The corresponding convergence trajectories compiled in \Cref{fig:cuboid_problem_convergence} demonstrate that the proposed PSFD framework exhibits an accelerated convergence profile compared to the conventional FDFD method, delivering a performance advantage with respect to both the total iteration count and the absolute computational execution time.

%
% FIGURE ON THE COMPARISON OF FDFD AND PSFD SOLVERS
%

Both baseline solvers in \Cref{fig:cuboid_problem_solved} account for the presence of the substrate stack by explicitly incorporating the entire multilayer region within the computational domain. This brute-force windowing approach becomes computationally prohibitive as the number of layers increases (e.g., more than the $20$ bilayers considered in the present example), which represents a standard industrial configuration for EUV masks. Furthermore, the strong field variations and standing-wave patterns localized inside the stack mandate a fine grid resolution along the vertical direction to resolve the sub-wavelength dynamics of the thinnest components (such as the $2.8\,$nm molybdenum layer within a standard $\mathrm{Mo/Si}$ bilayer pair).

We therefore implement the background field method via an outer fixed-point iteration loop based on \eqref{eq:field_recursive}. In this scheme, the reflected field generated by the stratified substrate is updated iteratively using the mask perturbation field computed from the preceding iteration. The boundary response at the interface is determined via the transfer-matrix method \cite{mackay2020}. Specifically, at an outer iteration step $i \in \mathbb{N}$, the formulation reduces to solving the following localized scattering problem for $\mathbf{E}_{\mathrm{mask}}^{(i)}$:
\begin{align}
    \nabla \times \nabla \times \mathbf{E}_{\mathrm{mask}}^{(i)} - \omega^2 \left(\varepsilon_{\mathrm{bg}} + \Delta \varepsilon_{\mathrm{mask}}\right) \mathbf{E}_{\mathrm{mask}}^{(i)} = \omega^2 \Delta \varepsilon_{\mathrm{mask}} \left(\mathbf{E}_{\mathrm{inc}} + \mathcal{R}\left[\mathbf{E}_{\mathrm{inc}} + \mathbf{E}_{\mathrm{mask}}^{(i-1)}\right]\right),
    \label{eq:outer_iteration_helmholtz}
\end{align}
where the reflection operator $\mathcal{R}$ acting on the boundary fields is defined as:
\begin{align}
    \mathcal{R}\left[\mathbf{E}_{\mathrm{inc}} + \mathbf{E}_{\mathrm{mask}}^{(i-1)}\right](z) = \mathrm{FFT}_{xy}^{-1} \left[ T_{\mathcal{R}} \cdot \mathrm{FFT}_{xy}\left[\mathbf{E}_{\mathrm{inc}}(z_{\mathrm{top}}) + \mathbf{E}_{\mathrm{mask}}^{(i-1)}(z_{\mathrm{top}})\right] \right] e^{i k_{z} (z - z_{\mathrm{top}})}, \quad z \ge z_{\mathrm{top}}.
    \label{eq:reflection_operator_fft}
\end{align}
Here, $z_{\mathrm{top}}$ denotes the interface between the layered stack and the absorber layer, $\mathrm{FFT}_{xy}$ represents the two-dimensional Fast Fourier Transform evaluated across the transverse plane, and $T_{\mathcal{R}}$ is the generalized reflection transfer matrix of the layered stack. This outer recursive sequence is executed continuously until the relative change in the reflected wavefield falls below a prescribed numerical convergence tolerance.

\begin{figure}[htbp]
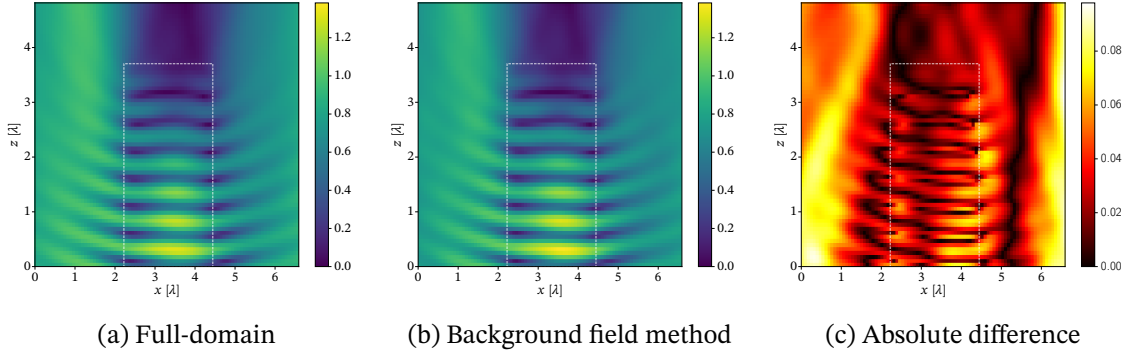

    \centering
    \subcaptionbox{Full-domain\label{fig:cuboid_problem_full}}{%
        \resizebox{0.32\textwidth}{!}{%
        \let\oldfontsize\fontsize
        \renewcommand{\fontsize}[2]{%
            \oldfontsize{\dimexpr1.4\dimexpr#1pt\relax\relax}{\dimexpr1.4\dimexpr#2pt\relax\relax}%
            \selectfont
        }
            \input{full_40.pgf}%
        }%
    }
    %\hspace{0.08\textwidth}
    \subcaptionbox{Background field method\label{fig:cuboid_problem_bg}}{%
        \resizebox{0.32\textwidth}{!}{%
        \let\oldfontsize\fontsize
        \renewcommand{\fontsize}[2]{%
            \oldfontsize{\dimexpr1.4\dimexpr#1pt\relax\relax}{\dimexpr1.4\dimexpr#2pt\relax\relax}%
            \selectfont
        }
            \input{bg_40.pgf}%
        }%
    }
    \subcaptionbox{Absolute difference\label{fig:cuboid_problem_bg_absdiff}}{%
        \resizebox{0.32\textwidth}{!}{%
        \let\oldfontsize\fontsize
        \renewcommand{\fontsize}[2]{%
            \oldfontsize{\dimexpr1.4\dimexpr#1pt\relax\relax}{\dimexpr1.4\dimexpr#2pt\relax\relax}%
            \selectfont
        }
            \input{full_bg_diff.pgf}%
        }%
    }
    \caption{Intensity distributions $\vert E_y\vert^2$ of the scattered electric fields on the $y=0$ plane, computed by (a) the full-domain PSFD and (b) the background field method together with (c) their absolute difference. The white dashed lines indicate the interfaces of the Ta absorber.}
    \label{fig:cuboid_problem_bg_solved}
\end{figure}

In \Cref{fig:cuboid_problem_bg_solved}, the electric field distribution computed using the proposed background field method is compared against the baseline full-domain PSFD solution obtained without the background decomposition. The discrepancy between the two wavefields, highlighted as an absolute difference map in \Cref{fig:cuboid_problem_bg_absdiff}, stems primarily from numerical artifacts inherent to the full-domain PSFD scheme, which directly resolves the multi-layered stack on the spatial grid resulting in discretization and phase-dispersion errors bypassed by the analytical reflection updates of our background formulation.

\begin{figure}[htbp]
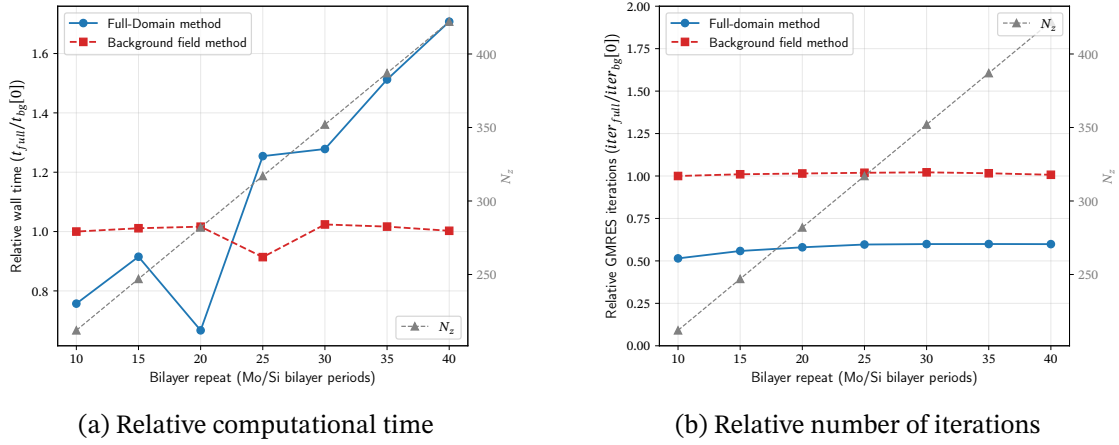

    \centering
    \subcaptionbox{Relative computational time\label{fig:full_bg_efficiency_wall_time}}{%
        \resizebox{0.45\textwidth}{!}{%
            \input{wall_time_vs_layers.pgf}%
        }%
    }
    \hspace{0.05\textwidth}
    \subcaptionbox{Relative number of iterations\label{fig:full_bg_efficiency_iterations}}{%
        \resizebox{0.45\textwidth}{!}{%
            \input{gmres_iterations_vs_layers.pgf}%
        }%
    }
    \caption{
        Comparison between the full-domain PSFD method and the proposed background field method for the test problem in \Cref{fig:cuboid_problem} across a varying number of multilayer stack layers. (a) Computational execution time, showing that the background field formulation remains invariant to the layer count while the full-domain method shows an escalating runtime penalty. (b) Total iteration count required to converge, illustrating that while the background field loop demands more iterations due to sequential recursive reflection updates.
    }
    \label{fig:full_bg_efficiency}
\end{figure}

As illustrated in \Cref{fig:full_bg_efficiency}, the proposed background field formulation demonstrates a clear superiority over the standard full-domain PSFD method in terms of total execution time (\Cref{fig:full_bg_efficiency_wall_time}) when applied to geometries with a high layer count (exceeding $25$ layers in this configuration). In general, the background field approach requires a larger number of total iterations to converge (\Cref{fig:full_bg_efficiency_iterations}) due to the fixed-point outer iteration loops mandating sequential updates of the reflection field. Crucially, however, because the active spatial grid size remains strictly bounded, the absolute computational runtime per solver step is uniform and invariant to the number of underlying substrate layers.

While the background field method guarantees a computational cost that is independent of the structural complexity of the underlying layered media, the remaining isolated scattering problem for the mask topography must still be resolved using an iterative Krylov solver whose convergence rate can be accelerated by introducing a preconditioning strategy \cite{saad2003,barrett1994}. To this end, we exploit the fact that the decoupled scattering formulation in \eqref{eq:field_recursive} is formally identical to a free-space scattering configuration defined over a homogeneous background medium with a uniform permittivity $\varepsilon_{\mathrm{bg}}$. Consequently, the global algebraic system matrix $\mathbf{A}$ can be split via $\mathbf{A} = \mathbf{A}_{\mathrm{bg}} - \mathbf{V}_{\mathrm{mask}}$, where $\mathbf{A}_{\mathrm{bg}}$ represents the discrete operator corresponding to the homogeneous background medium, and $\mathbf{V}_{\mathrm{mask}}$ denotes the localized diagonal perturbation mapping the mask's complex scattering potential.

Given that $\mathbf{V}_{\mathrm{mask}}$ has a compactly bounded support in the spatial domain, the left preconditioner $\mathbf{A}_{\mathrm{bg}}^{-1}$ (assuming non-singularity) can be expected to significantly accelerate the convergence rate of the iterative subspace solver. It achieves this by clustering the eigenvalues of the preconditioned operator system matrix $\mathbf{A}_{\mathrm{bg}}^{-1} \mathbf{A}$ tightly around unity \cite{nevanlinna1993,greenbaum1997,liesen2012}. The left-preconditioned linear system is explicitly expressed as:
\begin{align}
    \big(\mathbf{I} - \mathbf{A}_{\mathrm{bg}}^{-1} \mathbf{V}_{\mathrm{mask}}\big) \mathbf{x} = \mathbf{A}_{\mathrm{bg}}^{-1} \mathbf{b},
    \label{eq:preconditioned_system}
\end{align}
where $\mathbf{I}$ denotes the discrete algebraic identity operator.

Although algebraic preconditioning via explicit application of $\mathbf{A}_{\mathrm{bg}}^{-1}$ is conceptually attractive, it is computationally prohibitive. In particular, because the perfectly matched layers (PMLs) possess position-dependent conductivity profiles, the background operator $\mathbf{A}_{\mathrm{bg}}$ loses its spatial translational invariance. Consequently, it can no longer be diagonalized into a simple scalar pole in the Fourier domain; instead, it forms a dense mode-coupling system in spectral coordinates. The loss of spectral sparsity precludes a closed-form algebraic representation for $\mathbf{A}_{\mathrm{bg}}^{-1}$, rendering direct domain-inversion steps unviable within an $\mathcal{O}(\mathcal{N}\log\mathcal{N})$ matrix-free pseudo-spectral pipeline.

This bottleneck can be elegantly bypassed by recognizing that the inverse background operator $\mathbf{A}_{\mathrm{bg}}^{-1}$ is the discrete representation of the continuous dyadic Green's function for the homogeneous background medium. Consequently, the analytic counterpart to the preconditioned differential equation is the Lippmann-Schwinger volume integral formulation of the scattering problem \cite{colton2020}. Rather than explicitly attempting to precondition the stiff differential system algebraically through a dense matrix inversion, we can directly construct a naturally preconditioned integral equation derived from the Lippmann-Schwinger identity:
\begin{align}
    \mathbf{E}_{\mathrm{mask}}(\mathbf{r}) - \omega^{2} \int_{V_{\mathrm{mask}}} \overline{\mathbf{G}}_{0}(\mathbf{r}, \mathbf{r}') \cdot \Delta \varepsilon_{\mathrm{mask}}(\mathbf{r}') \mathbf{E}_{\mathrm{mask}}(\mathbf{r}') \, d\mathbf{r}' = \mathbf{E}_{\mathrm{inc}}^{\mathrm{tot}}(\mathbf{r}),
    \label{eq:lippman_schwinger}
\end{align}
where $\overline{\mathbf{G}}_{0}$ represents the free-space dyadic Green's function of the homogeneous background medium. Here, the driving excitation field $\mathbf{E}_{\mathrm{inc}}^{\mathrm{tot}}(\mathbf{r})$ is the total effective incident field containing the analytical reflections from the stratified substrate layers, evaluated over the mask domain as:
\begin{align}
    \mathbf{E}_{\mathrm{inc}}^{\mathrm{tot}}(\mathbf{r}) = \omega^{2} \int_{V_{\mathrm{mask}}} \overline{\mathbf{G}}_{0}(\mathbf{r}, \mathbf{r}') \cdot \left[ \Delta \varepsilon_{\mathrm{mask}}(\mathbf{r}') \Big( \mathbf{E}_{\mathrm{inc}}(\mathbf{r}') + \mathcal{R}\big[\mathbf{E}_{\mathrm{inc}}(\mathbf{r}') + \mathbf{E}_{\mathrm{mask}}(\mathbf{r}')\big] \Big) \right] d\mathbf{r}'.
    \label{eq:effective_incident_field}
\end{align}

With the Lippmann-Schwinger volume integral formulation established in \eqref{eq:lippman_schwinger}, we can adopt the PSFD framework to discretize the integral equation, utilizing the discrete Fourier domain to evaluate the spatial convolutions with optimal efficiency. Although the forward Fourier transform of the continuous dyadic Green's function $\overline{\mathbf{G}}$ can be derived analytically, its standard free-space singularity at the background wave resonance $|\mathbf{k}| = k_{\mathrm{bg}} = \omega \sqrt{\varepsilon_{\mathrm{bg}}}$ introduces severe numerical instability when mapped onto a discrete grid.

To bypass the resonance singularity, we introduce a windowed (truncated) dyadic Green's function $\overline{\mathbf{G}}_{R}(\mathbf{r})$, defined following the compact regularization techniques of \cite{vainikko2000,vico2016,pham2019}:
\begin{align}
    \overline{\mathbf{G}}_{R}(\mathbf{r}) = 
    \begin{cases}
        \overline{\mathbf{G}}(\mathbf{r}), & |\mathbf{r}| < R \\
        0, & |\mathbf{r}| \ge R
    \end{cases}
\end{align}
By ensuring that the truncation window radius $R$ is chosen to be strictly larger than the bounding diameter of both the active mask features and the local computational region, convolutions evaluated with $\overline{\mathbf{G}}_{R}(\mathbf{r})$ remain mathematically identical to those evaluated with the infinite-domain kernel. Moreover, by virtue of the Paley-Wiener theorem, the compact spatial support of $\overline{\mathbf{G}}_{R}(\mathbf{r})$ guarantees that its continuous spectral transform is an entire function, rendering its grid projection smoother and numerically more stable than the raw transform of $\overline{\mathbf{G}}(\mathbf{r})$.

The Fourier transform of the windowed dyadic Green's function $\overline{\mathbf{G}}_{R}$ can be derived analytically by exploiting the relation between the dyadic Green's function $\overline{\mathbf{G}}$ and the scalar Green's function $g$ of the homogeneous background medium:
\begin{align}
    \overline{\mathbf{G}}(\mathbf{r}) = \left( \overline{\mathbf{I}} + \frac{1}{k_{\mathrm{bg}}^{2}} \nabla \otimes \nabla \right) g(\mathbf{r}), \quad g(\mathbf{r}) = \frac{e^{i k_{\mathrm{bg}} r}}{4 \pi r}.
    \label{eq:dyadic_scalar_relation}
\end{align}
Using the identity \eqref{eq:dyadic_scalar_relation}, the spectral representation of the windowed dyadic Green's function $\mathcal{F}\big[\overline{\mathbf{G}}_{R}\big](\mathbf{k})$ is given by the decomposition:
\begin{align}
    \mathcal{F}\big[\overline{\mathbf{G}}_{R}\big](\mathbf{k}) = A(k) \overline{\mathbf{I}} + B(k) \hat{\mathbf{k}} \otimes \hat{\mathbf{k}},
\end{align}
where $k = |\mathbf{k}|$, and the coefficients $A(k)$ and $B(k)$ are defined as follows:
\begin{subequations}
\begin{align}
    A(k) &= \begin{cases} 
    \mathcal{F}[g_{R}](0) + \frac{R^{2}}{3 k_{\mathrm{bg}}^{2}} g'(R), & k = 0, \\
    \mathcal{F}[g_{R}](k) + h(k; R), & k > 0, 
    \end{cases} \\
    B(k) &= \begin{cases} 
    0, & k = 0, \\
    -\mathcal{F}[g_{R}](k) - 3 h(k; R) - \frac{1}{k_{\mathrm{bg}}^{2}}, & k > 0, 
    \end{cases}
\end{align}
\end{subequations}
with the auxiliary function $h(k; R) = \frac{4 \pi R g'(R)}{k_{\mathrm{bg}}^{2}} \frac{j_{1}(k R)}{k R}$. The spectral components of the windowed scalar kernel $\mathcal{F}[g_{R}](k)$ are determined by the analytical transform:
\begin{align}
    \mathcal{F}[g_{R}](k) = \begin{cases} 
    \frac{i R}{2 k_{\mathrm{bg}}} - \frac{i}{2 k_{\mathrm{bg}}^{2}} e^{i k_{\mathrm{bg}} R} \sin(k_{\mathrm{bg}} R), & k = k_{\mathrm{bg}}, \\
    \frac{1}{k_{\mathrm{bg}}^{2}} \left[ e^{i k_{\mathrm{bg}} R} (1 - i R k_{\mathrm{bg}}) - 1 \right], & k = 0, \\
    \frac{1}{k_{\mathrm{bg}}^{2} - k^{2}} \left[ e^{i k_{\mathrm{bg}} R} \left( \cos(k R) - i \frac{k_{\mathrm{bg}}}{k} \sin(k R) \right) - 1 \right], & \text{otherwise}.
    \end{cases}
\end{align}

Having established the Lippmann-Schwinger integral formulation in \eqref{eq:lippman_schwinger}, we employ the PSFD method to construct the discrete algebraic system. The volumetric convolution is implemented via the spectral kernel $\mathcal{F}\left[\overline{\mathbf{G}}_{R}\right]$ and the three-dimensional Fast Fourier Transform ($\mathrm{FFT}_{xyz}$):
\begin{align}
    \mathbf{x} - \mathrm{FFT}_{xyz}^{-1} \Big( \mathcal{F}\left[\overline{\mathbf{G}}_{R}\right] \odot \mathrm{FFT}_{xyz}[\mathbf{V}_{\mathrm{mask}} \mathbf{x}] \Big) = \mathrm{FFT}_{xyz}^{-1} \Big( \mathcal{F}\left[\overline{\mathbf{G}}_{R}\right] \odot \mathrm{FFT}_{xyz} [\mathbf{b}_{\mathrm{inc}} + \mathbf{b}_{\mathrm{ref}}] \Big),
    \label{eq:num_lippmann_schwinger}
\end{align}
where $\odot$ denotes the element-wise dyadic multiplication in the spectral domain, and $\mathbf{b}_{\mathrm{inc}}$ and $\mathbf{b}_{\mathrm{ref}}$ represent the incident and reflected background field vectors, respectively. To ensure that the periodic convolution implemented by the FFT accurately mimics the linear non-periodic convolution of the integral equation, the discrete domain must be extended using zero-padding to a size encompassing the combined support of the scattering potential and the windowed Green's function \cite{vico2016,pham2019}. 

The discrete linear operator $\mathbf{L} = \mathbf{I} - \mathbf{G}*\mathbf{V}_{\mathrm{mask}}$ exhibits a spectrum highly clustered around unity, a property arising from the compactness and regularity of the windowed Green's function kernel $\overline{\mathbf{G}}_{R}$ combined with the compactly supported scattering potential $\mathbf{V}_{\mathrm{mask}}$ \cite{colton2020,kress2013}. This spectral clustering property ensures the rapid, robust convergence of the iterative Krylov solver for the discretized Lippmann-Schwinger system.

\begin{figure}[t]
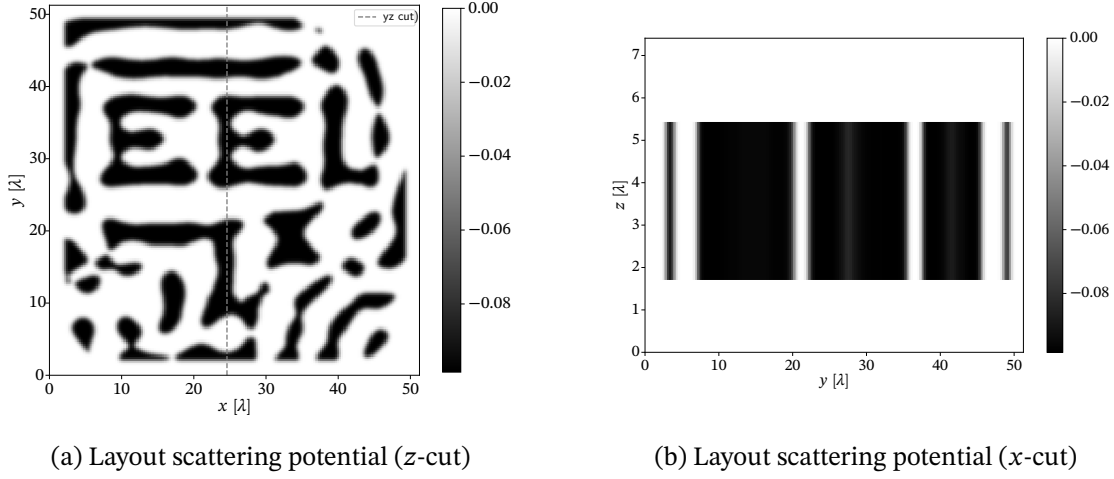

    \centering
    \subcaptionbox{Layout scattering potential ($z$-cut)\label{fig:layout_scattering_potential_z_cut}}{%
        \resizebox{0.45\textwidth}{!}{%
        \let\oldfontsize\fontsize
        \renewcommand{\fontsize}[2]{%
            \oldfontsize{\dimexpr1.4\dimexpr#1pt\relax\relax}{\dimexpr1.4\dimexpr#2pt\relax\relax}%
            \selectfont
        }
            \input{layout_xy_absorber_mid.pgf}%
        }%
    }
    \hspace{0.05\textwidth}
    \subcaptionbox{Layout scattering potential ($x$-cut)\label{fig:layout_scattering_potential_x_cut}}{%
        \resizebox{0.45\textwidth}{!}{%
        \let\oldfontsize\fontsize
        \renewcommand{\fontsize}[2]{%
            \oldfontsize{\dimexpr1.4\dimexpr#1pt\relax\relax}{\dimexpr1.4\dimexpr#2pt\relax\relax}%
            \selectfont
        }
            \input{layout_yz_center.pgf}%
        }%
        \vspace{0.38cm}
    }
    \caption{The sample mask structure taken from the LithoBench \cite{zheng2023} used for the demonstration of the Lippmann-Schwinger solver. The white dashed line in (a) indicates the $x$-cut to give (b).}
    \label{fig:layout_scattering_potential}
\end{figure}

\begin{figure}[t]
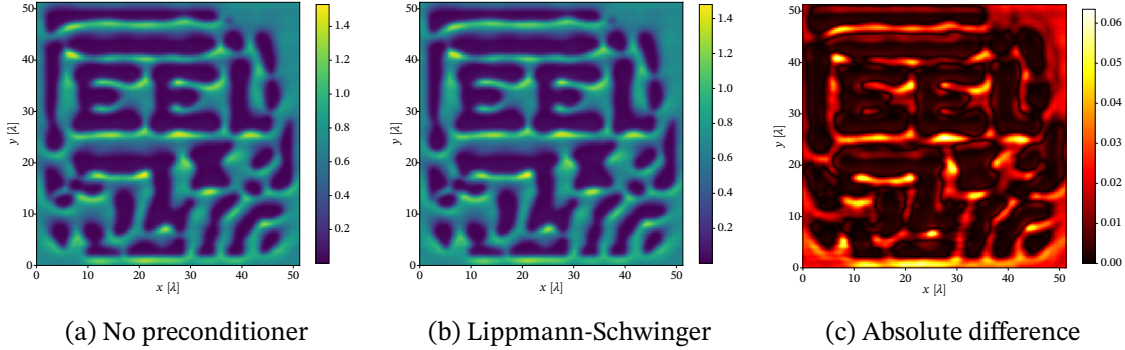

    \centering
    \subcaptionbox{No preconditioner\label{fig:layout_problem_damper}}{%
        \resizebox{0.32\textwidth}{!}{%
        \let\oldfontsize\fontsize
        \renewcommand{\fontsize}[2]{%
            \oldfontsize{\dimexpr1.4\dimexpr#1pt\relax\relax}{\dimexpr1.4\dimexpr#2pt\relax\relax}%
            \selectfont
        }
            \input{layout_xy_absorber_top_damper.pgf}%
        }%
    }
    \subcaptionbox{Lippmann-Schwinger\label{fig:layout_problem_ls}}{%
        \resizebox{0.32\textwidth}{!}{%
        \let\oldfontsize\fontsize
        \renewcommand{\fontsize}[2]{%
            \oldfontsize{\dimexpr1.4\dimexpr#1pt\relax\relax}{\dimexpr1.4\dimexpr#2pt\relax\relax}%
            \selectfont
        }
            \input{layout_xy_absorber_top_ls.pgf}%
        }%
    }
    \subcaptionbox{Absolute difference\label{fig:layout_problem_absdiff}}{%
        \resizebox{0.32\textwidth}{!}{%
        \let\oldfontsize\fontsize
        \renewcommand{\fontsize}[2]{%
            \oldfontsize{\dimexpr1.4\dimexpr#1pt\relax\relax}{\dimexpr1.4\dimexpr#2pt\relax\relax}%
            \selectfont
        }
            \input{layout_xy_absorber_top_diff.pgf}%
        }%
    }
    \caption{The total intensity distribution $\vert \mathbf{E}_{\mathrm{mask}} + \mathbf{E}_{\mathrm{bg}} \vert^2$ of the scattered field on top of the absorber, computed by (a) the background field method without preconditioning and (b) the Lippmann-Schwinger solver together with (c) their absolute difference.}
    \label{fig:layout_problem_solved}
\end{figure}

In \Cref{fig:layout_problem_solved}, we demonstrate the Lippmann-Schwinger (LS) solver by visualizing the total intensity distribution of the scattered field, which exhibits good agreement with results obtained by the original background field method. For this validation, we transition from the simple cuboid structure in \Cref{fig:cuboid_problem} to a more realistic mask layout (a $760 \times 760\,$nm$^{2}$ with $4\,$nm grid spacing) sampled from the LithoBench dataset \cite{zheng2023}. The geometry consists of a $50\,$nm-thick Ta absorber layer deposited atop a Mo/Si distributed Bragg reflector comprising $40$ bilayer pairs, as illustrated in \Cref{fig:layout_scattering_potential}. The system is excited by a $\mathrm{TE}$-polarized plane wave at a $13.5\,$nm wavelength with an incidence angle of $6^{\circ}$.

\begin{figure}[t]
    \centering{
        \resizebox{0.7\textwidth}{!}{%
        \let\oldfontsize\fontsize
        \renewcommand{\fontsize}[2]{%
            \oldfontsize{\dimexpr1.1\dimexpr#1pt\relax\relax}{\dimexpr1.1\dimexpr#2pt\relax\relax}%
            \selectfont
        }
            \input{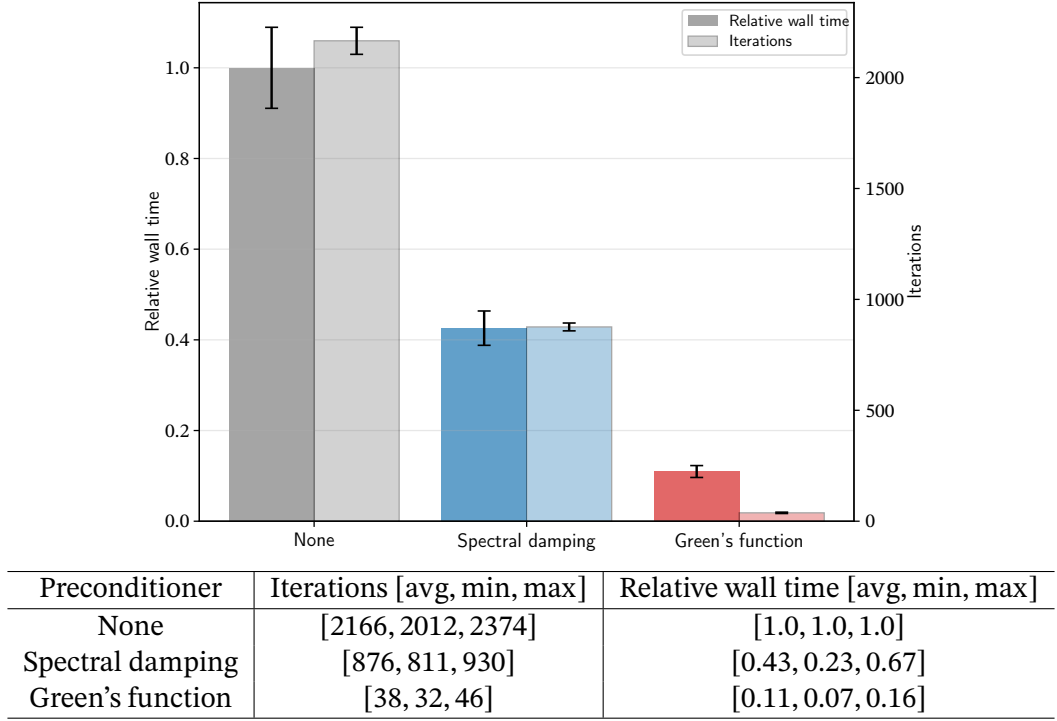}%

        }%
    \hfill}\\
    \begin{tabular}{c|c|c}
        \hline
        Preconditioner & Iterations $[\mathrm{avg}, \mathrm{min}, \mathrm{max}]$ & Relative wall time $[\mathrm{avg}, \mathrm{min}, \mathrm{max}]$ \\
        \hline
        None & $[2166, 2012, 2374]$ & $[1.0, 1.0, 1.0]$ \\
        Spectral damping & $[876, 811, 930]$ & $[0.43, 0.23, 0.67]$ \\
        Green's function & $[38, 32, 46]$ & $[0.11, 0.07, 0.16]$ \\
        \hline
    \end{tabular}
    \caption{The convergence behavior of the iterative solver with different preconditioners for $100$ randomly sampled layouts from the LithoBench \cite{zheng2023}. The Lippmann-Schwinger formula exhibits $23\times$ and $4\times$ speed up in the number of iterations and the computational time, respectively, compared to the spectral damping preconditioner.}
    \label{fig:ls_convergence_histogram}
\end{figure}

In \Cref{fig:ls_convergence_histogram}, we present a statistical benchmark evaluating the robustness of the proposed preconditioner against $100$ randomly sampled mask layouts from the LithoBench dataset \cite{zheng2023}. The simulation parameters for each layout are identical to the configuration described in \Cref{fig:layout_problem_solved}. To establish a rigorous performance baseline, we compare the Lippmann-Schwinger (LS) formulation against a standard isotropic spectral damping preconditioner defined by the operator $D(k) = \min(1, k_{\mathrm{bg}}^{2} / k^{2})$, which approximates the spectral attenuation of a scalar Green's function. The results indicate that the LS formulation achieves, on average, a $23\times$ reduction in iteration count and a $4\times$ improvement in total computational execution time compared to the spectral damping baseline, thereby confirming the superior effectiveness of the full-dyadic Green's function as an iterative solver preconditioner.

\section{Conclusion and outlook}
\label{sec:conclusion}
In this work, we have presented an accelerated numerical solver for EUV mask scattering problems based on the pseudo-spectral frequency-domain (PSFD) method. To overcome the computational bottleneck inherent in resolving large-scale stratified multilayer stacks, we proposed a background field decomposition method, which iteratively updates the reflection from the substrate using the transfer-matrix method. Furthermore, we introduced a Green's function-based preconditioning strategy, implemented by discretizing the Lippmann-Schwinger integral formulation within the PSFD framework. Numerical benchmarks confirm that the proposed Green's function preconditioner significantly accelerates the convergence of the iterative solver, offering a pronounced performance improvement over standard spectral damping techniques for realistic, industrial-scale mask layouts.

In its current formulation, the EUV scattering solver relies on the spatial disjointness between the mask topography and the stratified background media, which enables the efficient incorporation of layered reflections via the transfer-matrix method. However, practical lithographic configurations often present more complex geometric challenges. For instance, mask structures may be embedded within the absorber layer rather than deposited directly on the surface, and defect analysis often necessitates simulating anomalies that possess overlapping supports with the layered stack \cite{erdmann2017,thakare2022,windpassinger2003}. These configurations violate the assumption of spatial separation, necessitating a generalized scattering framework capable of handling non-disjoint material potentials.

In scenarios where the mask topography and the stratified substrate no longer satisfy the spatial separation condition, the background field contribution cannot be adequately modeled by the transfer-matrix method alone. Instead, one must reformulate the scattering problem by defining the scattering potential as the contrast between the local material distribution and the planarly layered background. This general problem is formally treated through volume integral formulations relying on the stratified-medium Green's function, a subject extensively addressed in the literature \cite{chew1995,cai2000,cai2013,yuan2025}. However, the efficient numerical evaluation and implementation of these layered Green's functions---particularly when managing a large number of thin-film layers---remains a computational challenge.

Although the Lippmann-Schwinger (LS) preconditioner formulated using the windowed free-space Green's function drastically accelerates the convergence of the iterative solver, its numerical implementation requires substantial zero-padding to compute linear convolutions via the FFT. This padding overhead can render the memory-inefficiency of the method for large-scale applications. Consequently, developing efficient numerical schemes to compute these convolutions without necessitating memory beyond the original computational domain remains a promising direction for future research. For instance, the Fast Multipole Method (FMM) \cite{yuan2025,rokhlin1985,greengard1987,carrier1988,ergul2014} provides a hierarchical algorithmic framework for computing these long-range interactions with $\mathcal{O}(N)$ or $\mathcal{O}(N \log N)$ complexity. By integrating an FMM-based acceleration into the PSFD framework, one could potentially achieve massive speedups in solving large-scale scattering problems while completely eliminating the memory overhead imposed by mandatory zero-padding.

Finally, the large-scale simulation of electromagnetic scattering from EUV masks extending several micrometers in the transverse dimension remains a formidable computational challenge. To address this, combining a domain decomposition method with the proposed PSFD solver represents a promising approach to efficiently handle extended computational domains. In such scenarios, a global spectral expansion may prove insufficient for imposing the non-trivial boundary conditions required at domain interfaces. Consequently, adopting a local spectral expansion strategy---such as the spectral element method \cite{boyd2001}---warrants investigation. The spectral element framework offers the flexibility to enforce complex interface boundary conditions while retaining the high-order accuracy characteristic of pseudo-spectral schemes, thereby facilitating a scalable solution for extended lithographic mask layouts.
\section*{Acknowledgments}

SL and DK express gratitude to Pervaiz Kareem, Hazem Mesilhy and Xuelong Shi for enlightening discussions during the initial stages of this work. This work has been enabled in part by the NanoIC pilot line. The acquisition and operation are jointly funded by the Chips Joint Undertaking, through the European Union's Digital Europe (101183266) and Horizon Europe programs\\(101183277), as well as by the participating states Belgium (Flanders), France, Germany, Finland, Ireland and Romania. For more information, visit \href{https://www.nanoic-project.eu}{nanoic-project.eu}.

\section*{Disclaimer}
Funded by the European Union. Views and opinions expressed are however those of the author(s) only and do not necessarily reflect those of the European Union or Chips Joint Undertaking. Neither the European Union nor the granting authority can be held responsible for them.
%\bibliographystyle{jhep}

%\endgroup

\begin{thebibliography}{99}

\bibitem{wong2001}
A.K.-K.~Wong, \emph{Resolution {{Enhancement Techniques}} in {{Optical Lithography}}}, SPIE (2001).

\bibitem{mack2007}
C.~Mack, \emph{Fundamental {{Principles}} of {{Optical Lithography}}: {{The Science}} of {{Microfabrication}}}, John Wiley \& Sons (2007).

\bibitem{ma2011}
X.~Ma and G.R.~Arce, \emph{Computational {{Lithography}}}, John Wiley \& Sons (2011).

\bibitem{erdmann2017}
A.~Erdmann, D.~Xu, P.~Evanschitzky, V.~Philipsen, V.~Luong and E.~Hendrickx, \emph{Characterization and mitigation of {{3D}} mask effects in extreme ultraviolet lithography}, \href{https://doi.org/10.1515/aot-2017-0019}{\emph{Adv. Opt. Technol.} {\bfseries 6} (2017) 187}.

\bibitem{erdmann2019}
A.~Erdmann, P.~Evanschitzky, G.~Bottiglieri, E.~van Setten and T.~Fliervoet, \emph{{{3D}} mask effects in high {{NA EUV}} imaging}, \href{https://doi.org/10.1117/12.2515678}{\emph{Proc. SPIE 10957, Extreme {{Ultraviolet}} ({{EUV}}) {{Lithography X}}}, {\bfseries109570Z} (2019)}.

\bibitem{levinson2020}
H.J.~Levinson, \emph{Extreme {{Ultraviolet Lithography}}}, SPIE (2020).

\bibitem{yee1966}
K.~Yee, \emph{Numerical solution of initial boundary value problems involving maxwell's equations in isotropic media}, \href{https://doi.org/10.1109/TAP.1966.1138693}{\emph{IEEE Trans. Antennas Propagat.} {\bfseries 14} (1966) 302}.

\bibitem{taflove1998}
A.~Taflove, \emph{Advances in {{Computational Electrodynamics}}: {{The Finite-difference Time-domain Method}}}, Artech House (1998).

\bibitem{moharam1981}
M.G.~Moharam and T.K.~Gaylord, \emph{Rigorous coupled-wave analysis of planar-grating diffraction}, \href{https://doi.org/10.1364/JOSA.71.000811}{\emph{J. Opt. Soc. Am., JOSA} {\bfseries 71} (1981) 811}.

\bibitem{moharam1995}
M.G.~Moharam, D.A.~Pommet, E.B.~Grann and T.K.~Gaylord, \emph{Stable implementation of the rigorous coupled-wave analysis for surface-relief gratings: Enhanced transmittance matrix approach}, \href{https://doi.org/10.1364/JOSAA.12.001077}{\emph{J. Opt. Soc. Am. A, JOSAA} {\bfseries 12} (1995) 1077}.

\bibitem{liu1997}
Q.H.~Liu, \emph{The {{PSTD}} algorithm: {{A}} time-domain method requiring only two cells per wavelength}, \href{https://doi.org/10.1002/(SICI)1098-2760(19970620)15:3<158::AID-MOP11>3.0.CO;2-3}{\emph{Microw. Opt. Technol. Lett.} {\bfseries 15} (1997) 158}.

\bibitem{abbott1982}
L.F.~Abbott, \emph{Introduction to the {{Background Field Method}}}, {\emph{Acta Phys. Polon. B} {\bfseries 13} (1982) 33}.

\bibitem{chew1995}
W.C.~Chew, \emph{Waves and {{Fields}} in {{Inhomogeneous Media}}}, Wiley-IEEE Press (1995).

\bibitem{chew2001}
W.~Chew, E.~Michielssen, J.M.~Song and J.M.~Jin, \emph{Fast and {{Efficient Algorithms}} in {{Computational Electromagnetics}}}, Artech House (2001).

\bibitem{colton2020}
D.~Colton and R.~Kress, \emph{Inverse {{Acoustic}} and {{Electromagnetic Scattering Theory}}}, Springer (2020).

\bibitem{born1999}
M.~Born and E.~Wolf, \emph{Principles of {{Optics}}: {{Electromagnetic Theory}} of {{Propagation}}, {{Interference}} and {{Diffraction}} of {{Light}}}, 7th ed., Cambridge University Press (2013).

\bibitem{jackson2009}
J.D.~Jackson, \emph{Classical {{Electrodynamics}}}, 3rd ed., Wiley (1998).

\bibitem{chew1985}
W.~Chew, \emph{Response of a source on top of a vertically stratified half-space}, \href{https://doi.org/10.1109/TAP.1985.1143634}{\emph{IEEE Transactions on Antennas and Propagation} {\bfseries 33} (1985) 649}.

\bibitem{mackay2020}
T.G.~Mackay and A.~Lakhtakia, \emph{The {{Transfer-Matrix Method}} in {{Electromagnetics}} and {{Optics}}}, Springer (2020).

\bibitem{pistor1998}
T.V.~Pistor, K.~Adam and A.~Neureuther, \emph{Rigorous simulation of mask corner effects in extreme ultraviolet lithography}, \href{https://doi.org/10.1116/1.590476}{\emph{J. Vac. Sci. Technol. B} {\bfseries 16} (1998) 3449}.

\bibitem{boyd2001}
J.P.~Boyd, \emph{Chebyshev and {{Fourier Spectral Methods}}}, 2nd ed., Dover Publications (2001).

\bibitem{fornberg1998}
B.~Fornberg, \emph{A {{Practical Guide}} to {{Pseudospectral Methods}}}, Cambridge University Press (1998).

\bibitem{liu1999}
Q.H.~Liu, \emph{Large-scale simulations of electromagnetic and acoustic measurements using the pseudospectral time-domain ({{PSTD}}) algorithm}, \href{https://doi.org/10.1109/36.752210}{\emph{IEEE Transactions on Geoscience and Remote Sensing} {\bfseries 37} (1999) 917}.

\bibitem{song2011}
D.~Song, L.~Yuan and Y.Y.~Lu, \emph{Fourier-matching pseudospectral modal method for diffraction gratings}, \href{https://doi.org/10.1364/JOSAA.28.000613}{\emph{J. Opt. Soc. Am. A, JOSAA} {\bfseries 28} (2011) 613}.

\bibitem{munro2014}
P.R.T.~Munro, D.~Engelke and D.D.~Sampson, \emph{A compact source condition for modelling focused fields using the pseudospectral time-domain method}, \href{https://doi.org/10.1364/OE.22.005599}{\emph{Opt. Express, OE} {\bfseries 22} (2014) 5599}.

\bibitem{berenger1994}
J.-P.~Berenger, \emph{A perfectly matched layer for the absorption of electromagnetic waves}, \href{https://doi.org/10.1006/jcph.1994.1159}{\emph{Journal of Computational Physics} {\bfseries 114} (1994) 185}.

\bibitem{makhotkin2021}
I.A.~Makhotkin, M.~Wu, V.~Soltwisch, F.~Scholze and V.~Philipsen, \emph{Refined extreme ultraviolet mask stack model}, \href{https://doi.org/10.1364/JOSAA.416235}{\emph{J. Opt. Soc. Am. A, JOSAA} {\bfseries 38} (2021) 498}.

\bibitem{saad1986}
Y.~Saad and M.H.~Schultz, \emph{{{GMRES}}: {{A Generalized Minimal Residual Algorithm}} for {{Solving Nonsymmetric Linear Systems}}}, \href{https://doi.org/10.1137/0907058}{\emph{SIAM J. Sci. and Stat. Comput.} {\bfseries 7} (1986) 856}.

\bibitem{saad2003}
Y.~Saad, \emph{Iterative {{Methods}} for {{Sparse Linear Systems}}}, 2nd ed., SIAM (2003).

\bibitem{barrett1994}
R.~Barrett, M.~Berry, T.F.~Chan, J.~Demmel, J.~Donato, J.~Dongarra et~al., \emph{Templates for the {{Solution}} of {{Linear Systems}}: {{Building Blocks}} for {{Iterative Methods}}}, SIAM (1994).

\bibitem{nevanlinna1993}
O.~Nevanlinna, \emph{Convergence of {{Iterations}} for {{Linear Equations}}}, Birkh\"auser, Basel (1993).

\bibitem{greenbaum1997}
A.~Greenbaum, \emph{Iterative {{Methods}} for {{Solving Linear Systems}}}, SIAM (1997).

\bibitem{liesen2012}
J.~Liesen and Z.~Strakos, \emph{Krylov {{Subspace Methods}}: {{Principles}} and {{Analysis}}}, Oxford University Press (2012).

\bibitem{vainikko2000}
G.~Vainikko, \emph{Fast {{Solvers}} of the {{Lippmann-Schwinger Equation}}},  in R.P.~Gilbert, J.~Kajiwara, Y.S.~Xu and R.P.~Gilbert (eds.) \emph{Direct and {{Inverse Problems}} of {{Mathematical Physics}}}, vol.~5, Springer (2000).

\bibitem{vico2016}
F.~Vico, L.~Greengard and M.~Ferrando, \emph{Fast convolution with free-space {{Green}}'s functions}, \href{https://doi.org/10.1016/j.jcp.2016.07.028}{\emph{J. Comput. Phys.} {\bfseries 323} (2016) 191} [\href{https://arxiv.org/abs/1604.03155}{{\ttfamily 1604.03155}}].

\bibitem{pham2019}
T.-a.~Pham, E.~Soubies, A.~Ayoub, J.~Lim, D.~Psaltis and M.~Unser, \emph{Three-{{Dimensional Optical Diffraction Tomography}} with {{Lippmann-Schwinger Model}}}, \href{https://doi.org/10.1109/TCI.2020.2969070}{\emph{{IEEE} {Transactions on Computational Imaging}}, {\bfseries 6} (2020) 727}.

\bibitem{kress2013}
R.~Kress, \emph{Linear {{Integral Equations}}}, 2nd ed., Springer (1999).

\bibitem{zheng2023}
S.~Zheng, H.~Yang, B.~Zhu, B.~Yu and M.~Wong, \emph{{{LithoBench}}: {{Benchmarking AI Computational Lithography}} for {{Semiconductor Manufacturing}}}, \href{https://proceedings.neurips.cc/paper_files/paper/2023/hash/604b9fa9e1c16284e6517d923cf9ff20-Abstract-Datasets_and_Benchmarks.html}{\emph{Advances in Neural Information Processing Systems} {\bfseries 36} (2023) 30243}.

\bibitem{thakare2022}
D.~Thakare, A.~Delabie and V.~Philipsen, \emph{Optimizing {{EUV}} imaging metrics as a function of absorber thickness and illumination source: Simulation case study of {{Ta-Co}} alloy}, \href{https://doi.org/10.1117/12.2640098}{\emph{Proc. SPIE 12472, 37th {{European Mask}} and {{Lithography Conference}}}, {\bfseries 124720A} (2022)}.

\bibitem{windpassinger2003}
R.~Windpassinger, N.~Rosenkranz, T.~Scherubl, P.~Evanschitzky, A.~Erdmann and A.~Zibold, \emph{{{EUV}} mask simulation for {{AIMS}}}, \href{https://doi.org/10.1117/12.518053}{\emph{Proc. SPIE 5256, 23rd Annual BACUS Symposium on Photomask Technology} (2003)}.

\bibitem{cai2000}
W.~Cai and T.~Yu, \emph{Fast {{Calculations}} of {{Dyadic Green}}'s {{Functions}} for {{Electromagnetic Scattering}} in a {{Multilayered Medium}}}, \href{https://doi.org/10.1006/jcph.2000.6583}{\emph{J. Comput. Phys.} {\bfseries 165} (2000) 1}.

\bibitem{cai2013}
W.~Cai, \emph{Computational {{Methods}} for {{Electromagnetic Phenomena}}: {{Electrostatics}} in {{Solvation}}, {{Scattering}}, and {{Electron Transport}}}, Cambridge University Press (2013).

\bibitem{yuan2025}
H.~Yuan, B.~Wang, W.~Zhang and W.~Cai, \emph{Fast {{Multipole Method}} for {{Maxwell}}'s {{Equations}} in {{Layered Media}}}, \href{https://arxiv.org/abs/2507.18491}{arXiv:2507.18491[math.NA]} (2025).

\bibitem{rokhlin1985}
V.~Rokhlin, \emph{Rapid solution of integral equations of classical potential theory}, \href{https://doi.org/10.1016/0021-9991(85)90002-6}{\emph{J. Comput. Phys.} {\bfseries 60} (1985) 187}.

\bibitem{greengard1987}
L.~Greengard and V.~Rokhlin, \emph{A fast algorithm for particle simulations}, \href{https://doi.org/10.1016/0021-9991(87)90140-9}{\emph{J. Comput. Phys.} {\bfseries 73} (1987) 325}.

\bibitem{carrier1988}
J.~Carrier, L.~Greengard and V.~Rokhlin, \emph{A {{Fast Adaptive Multipole Algorithm}} for {{Particle Simulations}}}, \href{https://doi.org/10.1137/0909044}{\emph{SIAM J. Sci. and Stat. Comput.} {\bfseries 9} (1988) 669}.

\bibitem{ergul2014}
O.~Ergul and L.~Gurel, \emph{The {{Multilevel Fast Multipole Algorithm}} ({{MLFMA}}) for {{Solving Large-Scale Computational Electromagnetics Problems}}}, John Wiley \& Sons (2014).

\end{thebibliography}
\end{document}